\documentclass{aastex631}

\def\arcsa#1#2{$#1^{\prime\prime}_{^\textrm{.}}#2$}

\shorttitle{I18043-FLASHING ATCA}
\shortauthors{Uscanga et al.}
\graphicspath{{./}{figures/}}
\def\sun{\hbox{$_{\odot}$}}

\begin{document}

\title{Evolution of the outflow in the water fountain source IRAS 18043$-$2116}
v6.31\footnote{Released on March, 2nd, 2023}

\correspondingauthor{Lucero Uscanga}
\email{l.uscanga@ugto.mx}

\author[0000-0002-2082-1370]{L. Uscanga}
\affiliation{Departamento de Astronom\'ia, Universidad de Guanajuato, A.P. 144, 36000 Guanajuato, Gto., Mexico \\}

\author[0000-0002-0880-0091]{H. Imai}
\affiliation{Center for General Education, Institute for Comprehensive Education, 
Kagoshima University \\
1-21-30 Korimoto, Kagoshima 890-0065, Japan}

\author[0000-0002-7065-542X]{J. F. G\'omez}
\affiliation{Instituto de Astrof\'isica de Andaluc\'ia, CSIC, Glorieta de la Astronom\'ia s/n, E-18008 Granada, Spain}

\author[0000-0002-2149-2660]{D. Tafoya}
\affiliation{Department of Space, Earth and Environment, Chalmers University of Technology, Onsala
Space Observatory, 439 92 Onsala, Sweden}

\author[0000-0002-5526-990X]{G. Orosz}
\affiliation{Joint Institute for VLBI ERIC, Oude Hoogeveensedijk 4, 7991 PD Dwingeloo, The Netherlands}

\author[0000-0001-9525-7981]{T. P. McCarthy}
\affiliation{School of Natural Sciences, University of Tasmania, Private Bag 37, Hobart, Tasmania 7001, Australia}

\author{Y. Hamae}
\affiliation{Department of Physics and Astronomy, Faculty of Science, 
Kagoshima University \\
1-21-35 Korimoto, Kagoshima 890-0065, Japan}

\author[0000-0003-4419-6132]{K. Amada}
\affiliation{Department of Physics and Astronomy, Graduate School of Science and Engineering, Kagoshima University \\
1-21-35 Korimoto, Kagoshima 890-0065, Japan}



\begin{abstract}
We present the spectral and spatial evolution of H$_2$O masers associated with the water fountain source IRAS 18043$-$2116 found in the observations with the Nobeyama 45~m telescope and the Australia Telescope Compact Array. We have found new highest velocity components of the H$_2$O masers (at the red-shifted side $V_{\rm LSR}\simeq376$~km~s$^{-1}$ and at the blue-shifted side $V_{\rm LSR}\simeq$  $-$165~km~s$^{-1}$), 
and the resulting velocity spread of $\simeq 540$ km s$^{-1}$
breaks the speed record of fast jets/outflows in this type of sources.
The locations of those components have offsets from the axis joining the two major maser clusters,
indicating a large opening angle of the outflow ($\sim60\arcdeg$). The evolution of the maser cluster separation of $\sim$2.9 mas yr$^{-1}$ and the compact ($\sim$\arcsa{0}{2}) CO emission source mapped with the Atacama Large Millimeter-submillimeter Array suggest a very short ($\sim$30 yr) timescale of the outflow. We also confirmed the increase in the flux density of the 22 GHz continuum source. The properties of the jet and the continuum sources and their possible evolution 
in the transition to the planetary nebula phase are further discussed.
\end{abstract}


%

\section{Introduction} \label{sec:intro}

Water fountains (WFs) are evolved stars, mostly in the post-asymptotic giant branch (post-AGB) phase that show H$_2$O maser emission tracing high-velocity collimated jets when they are observed at high-angular resolution (see \citealt{ima07, des12}; for reviews). The velocity spread in their H$_2$O maser spectra is typically $>$ 50 km s$^{-1}$, and can be as large as $\simeq$ 500 km s$^{-1}$ \citep{gom11}. Their H$_2$O maser emission thus traces significantly faster motions than the typical expansion velocities of circumstellar envelopes (CSEs) during the AGB phase (10--30 km s$^{-1}$; \citealt{sev97}).
The WFs identified so far seem to be in the short transition phase between the AGB and the planetary nebula (PN) phases of low-mass stars (initial stellar masses $\lesssim$\,4 $M\sun$; \citealt{kho21}). The short dynamical ages of the maser jets (5$-$100 yr; \citealt{ima07,taf20}) may indicate that WFs represent one of the first manifestations of collimated mass-loss in evolved stars.  
It is believed that jets carving the CSEs play an essential role in the shaping process of the PNe \citep{sah98,taf20}. Interestingly, the spatio-kinematics of WFs show a wide variety, from clear collimation and bipolarity to more complicated patterns such as multiple arc-shaped patterns \citep{oro19}, sometimes with additional low-velocity components 
that may be associated with the relics of slowly expanding circumstellar material, probably ejected during the AGB phase
\citep{ima13}.
A rapid deceleration of a fast WF jet may show up in the systematic velocity drifts of individual peaks in the maser spectrum. One such example is IRAS 18113$-$2503, in which the jet would have been launched at an initial velocity of 800 km s$^{-1}$ and decelerated to 200$-$300 km s$^{-1}$ within 20$-$30 years \citep{oro19}. 
Such deceleration is also expected for the jet in W43A, with more direct evidence \citep{taf20}. Thus single-dish and interferometric observations of H$_2$O masers in these sources are able to trace the evolution of the collimated jet or even measure directly the growth of the WF outflow.

IRAS 18043$-$2116 (hereafter abbreviated as I18043) was first reported as a WF by \citet{dea07} with H$_2$O maser emission, which covered a velocity range of 210 km s$^{-1}$ ($+$185 to $-$20 km s$^{-1}$) around the stellar systemic velocity of $V_{\rm LSR}\simeq$87.0~km~s$^{-1}$~km~s$^{-1}$ (\citealt{dea04}). 
Later \citet{wal09} reported H$_2$O maser emission over a velocity spread of nearly 400 km s$^{-1}$ ($+$289 to $-$109 km s$^{-1}$). 
More recently, \citet{per17} 
reported maser emission spread over a more red-shifted velocity range from 
$+$349 to $-$46 km s$^{-1}$.
Therefore, this WF may host the second fastest jet identified in the H$_2$O maser spectra just after IRAS 18113$-$2503 \citep{gom11}. 
The distance to I18043 is estimated to be $\sim$11~kpc using the Revised Kinematic Distance Calculator \citep{rei14} on the basis of the systemic velocity of this source ($V_{\rm LSR}\simeq$87.0 km~s$^{-1}$, see Section \ref{sec:results_ALMA} for more details).
This calculator is based on a Galactic model that is best suited for star-forming regions and therefore, its application to evolved stars 
can cause some uncertainty in the estimated distance. 
However, WFs are located close to the Galactic plane and they are expected to follow the Galactic rotation \citep{imai07}.
Also, this distance is in good agreement with the value of 8.20$\pm$1.75 kpc calculated from modeling the spectral energy distribution of I18043 \citep{vic15}. 
 
Here we present new detections of the highest velocity components of the masers in I18043 yielded with the 45~m telescope of Nobeyama Radio Observatory (NRO) 
and with the Australia Telescope Compact Array (ATCA).
These detections broke the record of the top speed of the jet in I18043. 
We also present the follow-up mapping observations of the H$_2$O masers with ATCA to locate the new maser components.
The radio continuum emission associated with I18043 was also mapped. 
The whole structure of the outflow has been revealed by the observation of CO $J=2\rightarrow1$ emission with the Atacama Large Millimeter-submillimeter Array (ALMA).

Section \ref{sec:Observations} summarizes those NRO, ATCA, and ALMA observations. Section \ref{sec:Results} presents the results of the observations. Section \ref{sec:Discussion} provides the discussion on the evolution of the high velocity jet traced by H$_2$O maser and CO emission. The time variability in the continuum emission, which has also been confirmed in the present work, is also discussed.  

\section{Observations} \label{sec:Observations}

\subsection{NRO 45~m telescope} \label{sec:obs_NRO}
The single-dish monitoring observations of H$_2$O and SiO masers towards the WFs have been conducted in the FLASHING
``Finest Legacy Acquisitions of SiO- and H$_2$O-maser Ignitions by the Nobeyama Generation". FLASHING has been described in detail in \cite{ima20} and \cite{ama22}, including the specification and the record of the observations with the NRO 45~m telescope since 2018 December. The summary paper of FLASHING will be published in a separate paper. Here we focus on the FLASHING observations of H$_2$O (at the rest frequency of 22.235080 GHz)
masers in I18043 conducted during 2019 January -- 2020 April, including the epochs of the new discovery of the highest velocity components and those within one month of the ATCA observations described later. Each of the observations lasted for about one hour.

We used three spectral windows to obtain a total velocity coverage of
$\sim$1600~km~s$^{-1}$. Each spectral window had 2048 spectral channels, each with a velocity resolution of 0.41~km~s$^{-1}$. 
Data reduction was carried out using the \textbf{\footnote{\url{https://www.nro.nao.ac.jp/~nro45mrt/html/obs/newstar/}}}JavaNewstar package in a standard manner, namely, data integration in the time domain followed by spectral baseline fitting in the maser emission-free spectral channels. We adopt a conversion factor of 2.8~Jy~K$^{-1}$ to convert the antenna temperature to the flux density scale. 

\vspace{0.8cm}
\subsection{ATCA} \label{sec:obs_ATCA}

The ATCA observations presented in this paper were conducted on 2020 April 7 and 30,
2020 June 4, and 2021 March 30 within project C3361, which is a follow-up of the sources monitored in the FLASHING program
in order to increase the temporal sampling of spectra, and to obtain information on the evolution of the spatial distribution of the H$_2$O
masers. 
Two independent spectral windows were observed, centered at 18 and 22 GHz, with a bandwidth of 2 GHz each, obtaining full linear polarization products. We set up the Compact Array Broadband Backend (CABB) in its 64M-32k mode, which samples each of these spectral windows in 32 broadband channels per polarization with a coarse spectral resolution of 64 MHz, for radio continuum observations. Several of the broadband channels were further zoomed in to observe the H$_2$O maser line with a finer spectral resolution of 0.42 km~s$^{-1}$. The total velocity coverage for the line was 1727 km~s$^{-1}$ on April 7, and 3453 km~s$^{-1}$ in the later three sessions.

Initial calibration was carried out using standard procedures in Miriad. The absolute flux density scale was calibrated with the source PKS 1934$-$638. PKS 0537$-$441 was also used as a bandpass calibrator for the first epoch, while PKS 1613$-$586 was used for the other three epochs. The sources IERS B1817$-$254, PMN J1832$-$2039, IERS B1817$-$254, and PMN J1755$-$2232
were used as complex-gain calibrators in each of the four epochs, respectively. Further processing (imaging, deconvolution, and self-calibration) was carried out with the Astronomical Image Processing System (AIPS). 
For the continuum data, we flagged out the broadband channels containing the maser line, to avoid contamination.
In the line data, the spectral channel with strongest maser emission 
was used for self-calibration as well as reference position for the maps.
The phase and amplitude corrections for self-calibration were obtained
at 10-second intervals (the integration time of each individual visibility). These solutions were then applied to both the whole spectral dataset and the broadband continuum data centered at 22 GHz. Since the line and continuum data  share the same calibration corrections, the $1\sigma$ relative positional accuracy among individual maser components and between maser and continuum emission is approximately given by the size of the synthesized beam divided by two times the signal-to-noise ratio (SNR) of the emission. 

We note that the beam is highly elongated along the north-south direction, which strongly worsens the positional accuracy in that direction. 
The synthesized beams were
\arcsa{2}{53}$\times$\arcsa{0}{34} at a position angle of $-$6.78\
and \arcsa{1}{42}$\times$\arcsa{0}{41} at a position angle of $-$1.27\arcdeg\,on 2020 April 30 and 2021 March 30, respectively. In these two sessions only, the uv coverage was good enough to map the maser emission while the spectra were obtained for all the four sessions previously mentioned. 
We roughly estimate a positional accuracy of $\sim$20 mas (1$\sigma$) along the minor axis of the beam (almost in east--west direction) for a component with an SNR$\sim$10.
The relative positions of the maser components with respect to the reference one were determined by Gaussian fitting using the AIPS task JMFIT, assuming a single Gaussian component in each velocity channel. This simple assumption looks valid for distinguishing clusters of the maser components described in Section \ref{sec:results_ATCA}. 

\subsection{ALMA} \label{sec:obs_ALMA}
I18043 was also observed with ALMA on 2019 January 18 (project code: 2018.1.00250.S). The details of these observations are presented by \cite{kho21}. The spectral setup covered a frequency range that contains emission of the CO\,($J$=2$\rightarrow$1) line at 230.538~GHz. In this paper we present only the spectrum and image of the CO\,($J$=2$\rightarrow$1) in order to estimate the systemic velocity of I18043 and compare the spatio-kinematical distribution of the masers with that of the molecular material. The integration time of $\sim$11~min yielded an RMS noise level of 3 mJy in the spectrum for a the velocity resolution of 1.5 km~s$^{-1}$, and a synthesized beam of \arcsa{1}{24}$\times$\arcsa{1}{18} with a position angle of $-$78\arcdeg. 
The relative position accuracy 
is $\sim$60 mas for a component with an SNR$\sim$10.

\section{Results} \label{sec:Results}

\subsection{FLASHING/ATCA spectra of masers}
\label{sec:results_NRO}

Figure \ref{fig:FLASHING} shows a time series of the spectra of H$_2$O masers of I18043 taken with the NRO 45~m telescope and ATCA during 2019 January 2 -- 2021 March 30. Because of the rapid variation in the spectral profile from one observation epoch to another, it is difficult to directly check the consistency of the flux density scales between the spectra of the NRO 45~m telescope and ATCA. However, the maser components around $V_{\rm LSR}\sim$82~km~s$^{-1}$ had flux densities of $\sim$4~Jy in both spectra of the two telescopes during 7--16 April 2020, suggesting that the flux density scales are roughly consistent with each other within $\sim$10\%. 

The FLASHING session on 2019 January 2 detected the red-shifted side of the new velocity components at $V_{\rm LSR}\simeq376$~km~s$^{-1}$, while the ATCA session on 2020 April 30 detected the blue-shifted side of the new velocity component at $V_{\rm LSR}\simeq$  $-$165~km~s$^{-1}$. These detections break the record 
of the velocity spread of the maser components in this source
($-100$~km~s$^{-1}$$>V_{\rm LSR}>350$~km~s$^{-1}$, \citealt{wal09,per17}); 
this suggests that the top speed of the jet in I18043 has to be revised.
Moreover, this velocity spread ($\simeq 540$ km s$^{-1}$) is the largest ever detected in the spectra of H$_2$O masers of WFs, beating the previously reported one for IRAS 18113$-$2503 ($\simeq$ 500 km s$^{-1}$, \citealt{gom11}).

It is difficult to determine whether these extreme velocity components may be decelerating, as suggested by  \citet{oro19}  for IRAS 18113$-$2503, 
due to their too short lifetimes within a few  months. However, their seemingly constant velocities within $\sim$1~km~s$^{-1}$ on this timescale may rule out such rapid deceleration (up to 10 km~s$^{-1}$month$^{-1}$).

\subsection{ATCA maps of H$_2$O masers and the continuum source}
\label{sec:results_ATCA}

Figure \ref{fig:ATCA-map} shows the maps of H$_2$O masers taken in I18043 with ATCA on 2020 April 30 and 2021 March 30. Although some of the maser components have relatively large positional uncertainties (up to 500 mas) in the declination offset, almost parallel to the elongation direction of the synthesized beam, the maser map resolves the structure of the bipolar outflow in the east--west direction. 
One can see two main clusters of maser components at (A)
$-75\leq V_{\rm LSR} \leq 83$~km~s$^{-1}$ and (B) $83\leq V_{\rm LSR} \leq 188$~km~s$^{-1}$, as well as another blue-shifted cluster (C) $-113\leq V_{\rm LSR} \leq -98$~km~s$^{-1}$. 
An additional red-shifted group  (D) $206\leq V_{\rm LSR} \leq 209$~km~s$^{-1}$ is also tentatively visible 
to the west of cluster B.
The most red-shifted component ($V_{\rm LSR}\sim$ 344~km~s$^{-1}$) also may be included in the group D 
while the most blue-shifted component ($V_{\rm LSR}\sim -$165~km~s$^{-1}$) can be included in the group C,
taking into account their positional uncertainties mentioned above. 

Figure \ref{fig:ATCA-map} also shows the location of the continuum source 
$(X, Y)=(15\pm 23, 165\pm 131)$ [mas] on 2020 April 30 and 
$(X, Y)=(-13\pm4,270\pm16)$ [mas] on 2021 March 30.  The continuum source had a flux density of $S_{\nu}=0.75\pm0.05$~mJy on 2020 April 30 and $S_{\nu}=0.97\pm0.09$~mJy on 2021 March 30. The location of the continuum source, detected at ${\rm SNR}\simeq36$, also may have a large positional uncertainty related to the synthesized beam shape. Nevertheless, it is clear that the continuum source is located between
the blue- and red-shifted H$_2$O maser clusters. 

These ATCA maps taken in 2020--2021 can be directly compared with Fig.\ 3 of \citet{per17} taken in 2013. Although it is difficult to quantitatively compare the separations of the maser clusters in a time span of $\sim$7 years, one can see that the separation between the clusters A and B looks consistent. 
Taking into account the stellar systemic velocity of 
$V_{\rm LSR}\simeq$~87~km~s$^{-1}$ (\citealt{dea04}), 
the clusters A and B may form a pair with spectral and spatial symmetry with respect to the central star. Regarding the stellar systemic velocity, we revised its value using CO emission detected with ALMA, see Section \ref{sec:results_ALMA}. 

In the same manner, the cluster C and the group D will form another symmetric pair. However, the cluster C and the group D in 2020--2021 have large uncertainties in their locations and distributions. The group D in 2020--2021 may be coincident to the cluster B in 2013 \citep{per17}, while the cluster C has been newly visible $\sim$50~mas east of the cluster A.  

Figure \ref{fig:ATCA-map2} highlights the maser distributions during 2020--2021 in the east--west direction in which the relative maser positions were better determined. The separation between the two major clusters A and B is discussed in more detail in Section \ref{sec:Discussion}. 

\subsection{ALMA maps of CO lines} \label{sec:results_ALMA}

Figure \ref{fig:ALMA_CO2-1_spectrum} shows the spectrum of the CO($J$=2$\rightarrow$1) emission observed with ALMA. The line consists of a strong central component and high-velocity wings, which are indicated in the left panel of Figure \ref{fig:ALMA_CO2-1_spectrum}. The high-velocity wings extend over a velocity range $>$200~km~s$^{-1}$, although the blue-shifted wing exhibits contamination of foreground gas that probably belongs to molecular clouds along the line of sight. From the velocity of line peak of the central component, a systemic velocity of $V_{\rm LSR, sys}=87\pm1$~km~s$^{-1}$ is derived for I18043. This value is compatible with the systemic velocity obtained by \cite{dea04} using OH maser observations.

The spatial distribution of the CO($J$=2$\rightarrow$1) and 230~GHz continuum emission is shown in Figure \ref{fig:ALMA_CO2-1_map}. 
There exists an offset of the brightness centroids between the blue- and red-shifted components ($<$\arcsa{0}{2}) and spatially unresolved brightness distributions of these components well support the far distance to this source and the compactness of the jet structure ($\lesssim$\, 11 kpc$\times$\arcsa{0}{2} $\sim$2200 au). 

The locus of the CO($J$=2$\rightarrow$1) emission peak in each bin with a velocity width of 6 km s$^{-1}$ is presented in Figure \ref{fig:ALMA_pos_PV}. There is a clear velocity gradient in R.A. offset with respect to the 230 GHz continuum peak as shown in the right panel of Figure \ref{fig:ALMA_pos_PV}.

\section{Discussion} \label{sec:Discussion}
\subsection{Characteristics of the outflow in I18043 from the H$_2$O maser and CO line emissions} \label{sec:discussion_H2O_CO}
We measured the separation between the two main clusters of H$_2$O masers, blue- and red-shifted with respect to the systemic velocity $\simeq$87~km~s$^{-1}$. 
This separation of the two clusters was $\sim$45$\pm$5~mas on 2008 July 12 \citep{wal09}. 
On the other hand, it has changed to 
$\sim$$74.1\pm0.4$ mas
and $87.6\pm0.5$~mas on 2020 April 30 and 2021 March 30, respectively, as shown in Figure \ref{fig:ATCA-map2}. Note that the uncertainties in these separations are statistical.
This result may indicate a possible expansion of the cluster separation at a rate of 
$\sim$ 2.9$\pm$0.5 mas yr$^{-1}$, namely a growth of a bipolar outflow hosting the two clusters.

We also note that the angular separations of the red- and blue-shifted groups of H$_2$O masers in I18043, $\sim$74 and 88~mas, correspond to a linear length of $\sim$800 and 950~au, respectively. They are quite similar to the separation of the red- and blue-shifted groups of masers in IRAS 16552$-$3050 ($\sim$700 $\pm$ 100~au; \citealt{sua08,vic15}).
On the other hand, the velocity spreads are rather different between these two WFs
($\sim$540 and $\sim$170 km s$^{-1}$, \citealt{sua08}). The origin of the difference in the velocity spreads (inclination of the jet major axis, stellar luminosity,  evolution phase) will be further investigated.

In a previous analysis of VLBA data from 2008 October 26 and 2009 January 25 in I18043, \cite{oro17} was able to distinguish two distinct arc-shaped groups of masers, blue- and red-shifted with respect to the systemic velocity of $\sim$87~km~s$^{-1}$. The blue-shifted maser emission was within a velocity range from $-$111 km s$^{-1}$ to 78 km s$^{-1}$ and the red-shifted maser emission was within a velocity range from 94 km s$^{-1}$ to 176 km s$^{-1}$. Comparing these results with those by \cite{wal09}, it seems that the high-velocity components detected previously at velocities higher than 200 km s$^{-1}$ were not included in the VLBA epochs, with a velocity coverage up to 227 km s$^{-1}$ \citep{day11}. 
In our recent observations, we were able to detect higher velocity components than those previously reported,
in both blue and red-shifted sides of the spectra
at velocities up to $V_{\rm LSR}\simeq-$165 km s$^{-1}$ and $\simeq$ 376 km s$^{-1}$, respectively.
This velocity spread is the largest ever detected in a WF $\simeq$ 540 km s$^{-1}$.
The emergence of the new highest velocity components at the outer outflow lobes may indicate a rapid growth of the outflow triggered by an increase in the maximum outflow velocity. 

From two epochs of the VLBA data, 2008 October 26 to 2009 January 25; \cite{oro17} estimated roughly the expansion rate of the outflow to be $\sim$1 mas yr$^{-1}$ with an average separation of the two maser clusters of $\sim$63 mas. If we extrapolate the expansion rate of the outflow from 2009 to 2020 the separation of the two clusters would be $\sim$74 mas, which is consistent with the results of the ATCA observations. 
Thus the dynamical age of the outflow is estimated to be $t_{\mathrm{jet}}\lesssim$ 30 yr, 
using the expansion rate and the separation of the two clusters from the VLBA data. \citet{oro17} used the full velocity spread of the H$_2$O masers seen in their VLBA maps ($+$176 to $-$111 km s$^{-1}$) and derived the inclination angle of the jet's major axis to be 
$i_{\mathrm{jet}}\sim 75^\circ$ with respect to the plane of the sky. Using this inclination angle,
we estimate a 3D expansion velocity for the red-shifted side of  $V_{\mathrm{jet,3D}}\sim170$ or 300 km s$^{-1}$, 
when assuming mean proper motions of $\sim$1 and 2.9 mas yr$^{-1}$, respectively.
Considering the most red-shifted components observed with ATCA of $V_{\rm LSR}\sim$284 and $344$ km s$^{-1}$, respectively.

The  CO($J$=2$\rightarrow$1) emission consists of a spatially extended low-velocity component, shown as green contours in Figure \ref{fig:ALMA_CO2-1_map}, and a compact high-velocity component, shown as blue and red contours. The green contours represent the velocity-integrated emission in the velocity range: $+65$ to $+$110~km~s$^{-1}$, and the blue and red contours represent the velocity-integrated emission in the velocity ranges:$-$125 to $+65$~km~s$^{-1}$ and 
$+110$ to $+300$~km~s$^{-1}$, respectively. The peaks of the blue- and red-shifted components have an 
offset in the East-West direction by approximately 0$\rlap{.}^{\prime\prime}$2 with the blue-shifted emission located toward the East. Thus, the velocity gradient seen in the CO emission is in agreement with the velocity gradient of the blue- and red-shifted masers.

The right panel of Figure \ref{fig:ALMA_pos_PV} shows the  locus of the CO($J$=2$\rightarrow$1) emission peak in each bin with a velocity width of 6 km s$^{-1}$. 
The observed dispersion of the locus in the R.A. offset  indicates a wide-angle of the outflow. 
We note that there exists a clear velocity gradient in the R. A. offset of the CO emission from $-$20 to 20~km~s$^{-1}$ with respect to the systemic velocity. This velocity gradient is roughly followed by the blue- and red-shifted CO emission, considering the uncertainties and a SNR of 10. 
 
The wanders of the peak positions in the blue- and red-shifted CO emission components, $V_{\rm offset}>$ 70 km s$^{-1}$ may be affected by the large positional uncertainty, 
while those in the range,  $V_{\rm offset} =10-70$~km~s$^{-1}$ are roughly consistent with the ones of H$_2$O masers. 
This suggests that the H$_2$O masers are associated with the CO outflow itself, or possibly the shock regions in the outflow.  
The absence of CO emission at the fastest velocity range $\vert V_{\rm offset}\vert>100$~km~s$^{-1}$, 
where the H$_2$O masers are visible,
is different from the case of W43A in which the H$_2$O masers are associated with the slower-moving entrained material rather than the faster jet \citep{taf20}. The  CO velocity wanders also indicate  
a large opening angle of the outflow, within which the locations of different velocity components are scattered due to an inhomogeneity of the outflow velocity.
Taking into account the direction coverage of the wandering velocity components with respect to the originating point of the outflow, a full opening angle of the outflow
is estimated to be $\sim$$60^\circ$. 

Also looking at 
the right panel of Figure \ref{fig:ATCA-map} in more detail,
the blue-shifted components at $V_{\rm LSR}$ $\sim-104$ to $-101$~km~s$^{-1}$ present an offset from the two major clusters of masers A and B. This separation also may imply the presence of a wide-angle outflow, indeed supported by the CO($J$=2$\rightarrow$1) observations. 

Figure 7 shows a schematic picture of the WF I18043, including the currently 
observational panorama discussed in the previous sections.
The continuum emission, close to the central stellar system, may correspond to an ionized area at the inner part of a biconical outflow.
Farther away from the central system, around several hundreds of au, thick biconical cavities surrounding the outflow are expected from the observed clusters of H$_2$O masers, 
which are likely excited in the cavities by shock interaction with the outflow. 
The blue- and red-shifted H$_2$O masers may be more amplified along the front and rear sides of the inner cavities, where a longer amplification path is expected closely to the line of sight.
Moreover, the CO emission is tracing an envelope of several thousands of au. 
The velocity wanders in the blue- and red-shifted components of the CO emission indicate a large opening angle of the biconical outflow (a full angle of $\sim$60\arcdeg\, in each component). This is also consistent with a large scatter of the maser distribution in the direction perpendicular to the major axis of the outflow. 

Apart from the H$_2$O maser emission detected at 22 GHz, this source also presents H$_2$O maser emission at 321 GHz. These maser components span a velocity range similar to that 
at 22 GHz, indicating that these H$_2$O masers at both frequency transitions probably coexist. The intensity of the submillimeter masers is comparable to that of the 22 GHz masers, implying the kinetic temperature of the region where the submillimeter masers originate is $T_k >1000$ K \citep{taf18}. 

\subsection{Characteristics of the radio continuum emission in I18043} \label{sec:discussion_radio_continuum}
The origin of the radio continuum emission in this source is also relevant to understand its evolution. \citet{per17} obtained the flux density data at four  frequency bands between 1.5 and 22 GHz. 
They reported spectral indices ($\alpha$, defined as $S_\nu \propto \nu^\alpha$) $<2$ between 1.5 and 5.5 GHz. Specifically, we estimated a value of $0.49\pm 0.09$ from their data between 1.5 and 5.5 GHz, 
and $\alpha = -0.2\pm 0.2$ between 5.5 and 22 GHz. These values are consistent with free-free radiation from a plasma, which ranges from $\alpha = 2$ (optically thick regime) to $-$0.1 (optically thin), with an optical depth decreasing at higher frequencies. As mentioned by these authors, the spectral indices should be taken with care, since these observations were not simultaneous: the data at 22 GHz were taken in 2015, 
two years 
after those at lower frequencies, and the radio continuum emission in evolved stars is known to be variable \citep{cer11}. \citet{per17} argued that if the radio continuum emission happened to fade with time, then the flux density at 22 GHz at the time of the low-frequency observations (2015) would have been lower, leading to a spectral index at high frequencies significantly lower than $-$0.1, which in its turn, would point out to a non-thermal contribution. Actually, a decreasing flux density has been suggested in itself as being a signature of non-thermal radio continuum emission \citep{sua15,cer17}. 
However, following this line of reasoning, our data at 22 GHz does not show any evidence for a non-thermal component, since the radio continuum emission at 22 GHz seems to be consistently increasing: while a flux density of $0.70\pm 0.10$ in 2013 July was obtained by \citet{per17}, we measured $S_{\nu}=0.75\pm0.05$~mJy on 2020 April 30 and $S_{\nu}=0.97\pm0.09$~mJy on 2021 March 30. This is a factor of 1.3 increase in one year in our observations, with an apparently higher increase rate in our observations than in the previous ones by
\citet{per17}. 

Considering then that the current radio continuum data is compatible with thermal free-free emission from electrons in a plasma, we note that the spectral index $\alpha=0.49\pm 0.09$ in the \cite{per17} data between 1.5 and 5.5 GHz indicates a partially optically thick emission. It is also close to the standard value of $n_e\propto r^{-2}$ expected in the case of an isotropic distribution of electrons with a radial variation of electron density of $n_e\propto r^{-2}$, where $r$ is the distance to the central star \citep{oln75,pan75}. This would be the case of an isotropic ionized wind, or a photoionized region at the beginning of the PN phase (since the previously ejected envelope 
would also have an $r^{-2}$ density dependence). Under the assumption of an isotropic wind, \cite{per17}  obtained a mass-loss rate of $\simeq 3\times 10^{-5}$ ($D/11$ kpc) $M_\odot$ yr$^{-1}$.

However, I18043 clearly shows the presence of a collimated jet traced by H$_2$O maser emission, therefore assuming an isotropic ionized wind may not be appropriate. 
In this sense, \cite{per17} mentioned that the mass-loss rate they obtained under this assumption is an upper limit to the actual value.
On the other hand, a collimated, ionized jet with $n_e\propto r^{-2}$ would also show a similar spectral index in its radio continuum emission \citep{rey86}, with a particular value depending on the jet geometry. In this model, a biconical jet (constant opening angle) would give $\alpha=0.6$, while lower or higher values correspond to a jet whose width decreases or increases with distance from the star, respectively. 
Here the jet half-width is given by $w\propto r^\epsilon$, where $\epsilon$ is related to the spectral index as $\alpha= 1.3 - 0.7/\epsilon$ following Reynold's model. The case of a standard spherical wind yields $\epsilon=1$ and $\alpha=0.6$. On the other hand, an isothermal, constant-velocity, fully ionized flow yields $0.5<\epsilon\leq1$ and $0.6\geq\alpha>-0.1$. 
For the case of I18043, $\alpha=0.49$ and $\epsilon\simeq 0.89$, therefore, the radio continuum emission is congruous with a fully-ionized collimated outflow. 

In an ionized jet, the flux density of the radio continuum emission can provide an estimate of the mass-loss rate. Using Equation 8 in \citet {ang18} and the parameters from the jet traced by H$_2$O masers,
 we estimate $\dot{M} \simeq 4\times 10^{-6}$ M$_\odot$ yr$^{-1}$ and $\simeq 6\times 10^{-6}$ M$_\odot$ yr$^{-1}$, considering the flux density at 5.5 GHz and the spectral index $\alpha=0.49$ measured by \cite{per17} in 2013, a jet velocity of 170 and 300 km s$^{-1}$, respectively. In this calculation we assumed an injection
opening angle of the jet equal to $60^\circ$,
a distance to the source of 11 kpc, an electron temperature of $10^4$ K, an inclination angle $75^\circ$ with respect to the plane of the sky, and a turnover frequency of $\simeq 5.5$ GHz. 
On other hand, we obtain the following values for the mass-loss rates from $2$
to $4\times 10^{-6}$  M$_\odot$ yr$^{-1}$, when they are derived from the flux density measurements at 22 GHz in 2020$-$2021, 
although these are likely to be lower limits, considering that the continuum emission at this frequency may be in the optically thin regime \citep{per17}. This would suggest that, if the continuum emission arises from a jet, its mass-loss is increasing. In any case, better estimation of the physical parameters would require an up-to-date measurement of the continuum spectrum of the source, with (nearly) simultaneous observations at a wide range of frequencies.
Even so these values of the mass-loss rates are not very accurate; they seem to be higher than those expected in models of post-AGB stars 
($\simeq 10^{-7}-10^{-8}$ M$_\odot$ yr$^{-1}$, \citealt{vas94}), although these models follow the evolution of a single star. 
Lately, observations and models suggest that the presence of collimated jets in WFs could be due to a binary system in the center,
leading to the formation of accretion disks, from which the collimated outflow can be launched. In particular, these objects seem to have gone through a common envelope phase \citep{kho21,gar21}, which greatly enhances mass-loss rates. 

An alternative explanation for the origin of the radio continuum emission is an onset of photoionization, at which this source is entering the PN phase. 
An increase in the flux density at radio wavelengths has
also been observed in several young PNe \citep{kwo81,kna95,chr98,gom05,taf09,cal22}. 
Such a flux density increase has been interpreted as an expansion of the ionized region. If this is the case, then an increase by a factor of 1.3 in flux density would correspond to an expansion by a factor of 1.14 in radius in one year, for an isotropic ionized region. 
We should consider that the size of the emitting region is unresolved in our observations, therefore we are not able to estimate other physical parameters such as the mass of the ionized region.
Furthermore, the scenario of a photoionized region around this source has also been discussed by \cite{per17}, but they concluded that a shock-ionized wind is more likely to explain the observed radio continuum emission. Either a shock- or radiative-ionized region around I18043, 
further observations would be necessary to identify the actual scenario. 

\section{Summary} \label{sec:Summary}
The spectral and spatial evolution of the H$_2$O masers in I18043 observed with Nobeyama 45m telescope and ATCA reveal new highest velocity components located at the outer outflow lobes (up to $\sim$ 950 au), suggesting a rapid growth of the outflow triggered by an increase in the maximum outflow velocity. 
This source presents the largest velocity spread ever detected in the spectra of H$_2$O masers of a WF $\simeq 540$ km s$^{-1}$.
The spatial distribution of the H$_2$O masers may indicate 
the presence of an outflow with a large opening angle $\sim$60$^\circ$ and a very short timescale $\sim$30 yr. 

\begin{acknowledgments}

The Nobeyama 45-m radio telescope is operated by Nobeyama Radio Observatory, a  branch of National Astronomical Observatory of Japan (NAOJ), National Institutes of Natural Sciences. The Australia Telescope Compact Array is part of the Australia Telescope National Facility (grid.421683.a) which is funded by the Australian Government for operation as a National Facility managed by CSIRO. We are thankful to the referee for his/her useful comments that helped us to improve this paper. LU acknowledges support from the University of Guanajuato (Mexico) grant ID CIIC 164/2022. HI and GO are supported by the JSPS KAKENHI Grant Number JP16H02167. 
JFG acknowledges support from grants PID2020-114461GB-I00 and CEX2021-001131-S, funded by MCIN/ AEI /10.13039/50110001103.
HI and JFG were supported by the Invitation Program for Foreign Researchers of the Japan Society for Promotion of Science (JSPS grant S14128). DT was supported by the ERC consolidator grant 614264. GO was supported by the Australian Research Council Discovery project DP180101061 of the Australian government, and the grants of CAS LCWR 2018-XBQNXZ-B-021 and National Key R\&D Program of China 2018YFA0404602. 

\end{acknowledgments}

\vspace{5mm}
\facilities{Nobeyama 45~m telescope, ATCA, ALMA}

\begin{figure*}
\includegraphics[width=18cm]{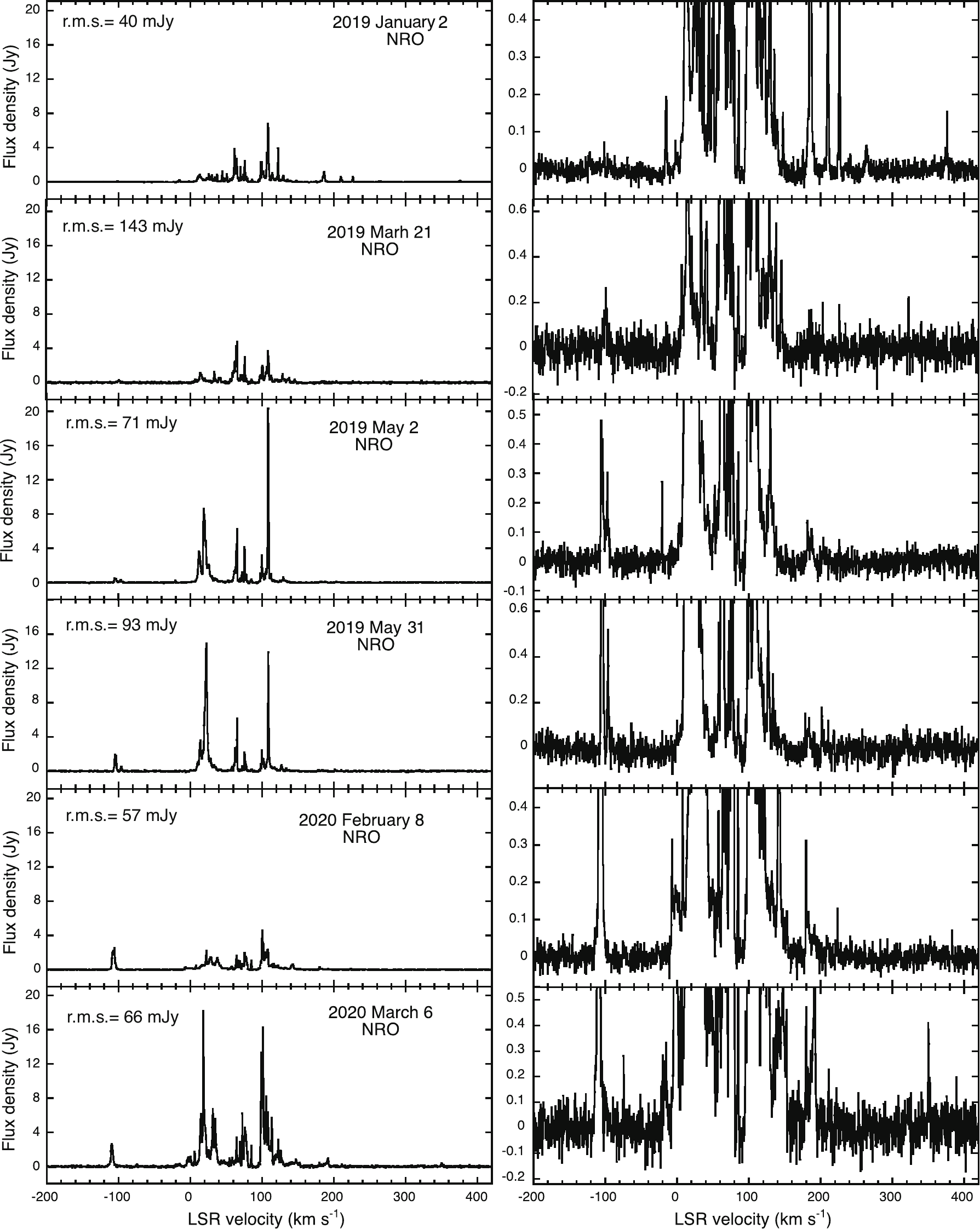} 
\caption{Spectra of H$_2$O masers in I18043 taken with ATCA and NRO telescopes during 2019 January 2--2021 March 30.
 The root-mean-square (RMS) noise levels are indicated in each spectrum.}
\label{fig:FLASHING}
\end{figure*}

\addtocounter{figure}{-1}
\begin{figure*}
\includegraphics[width=18cm]{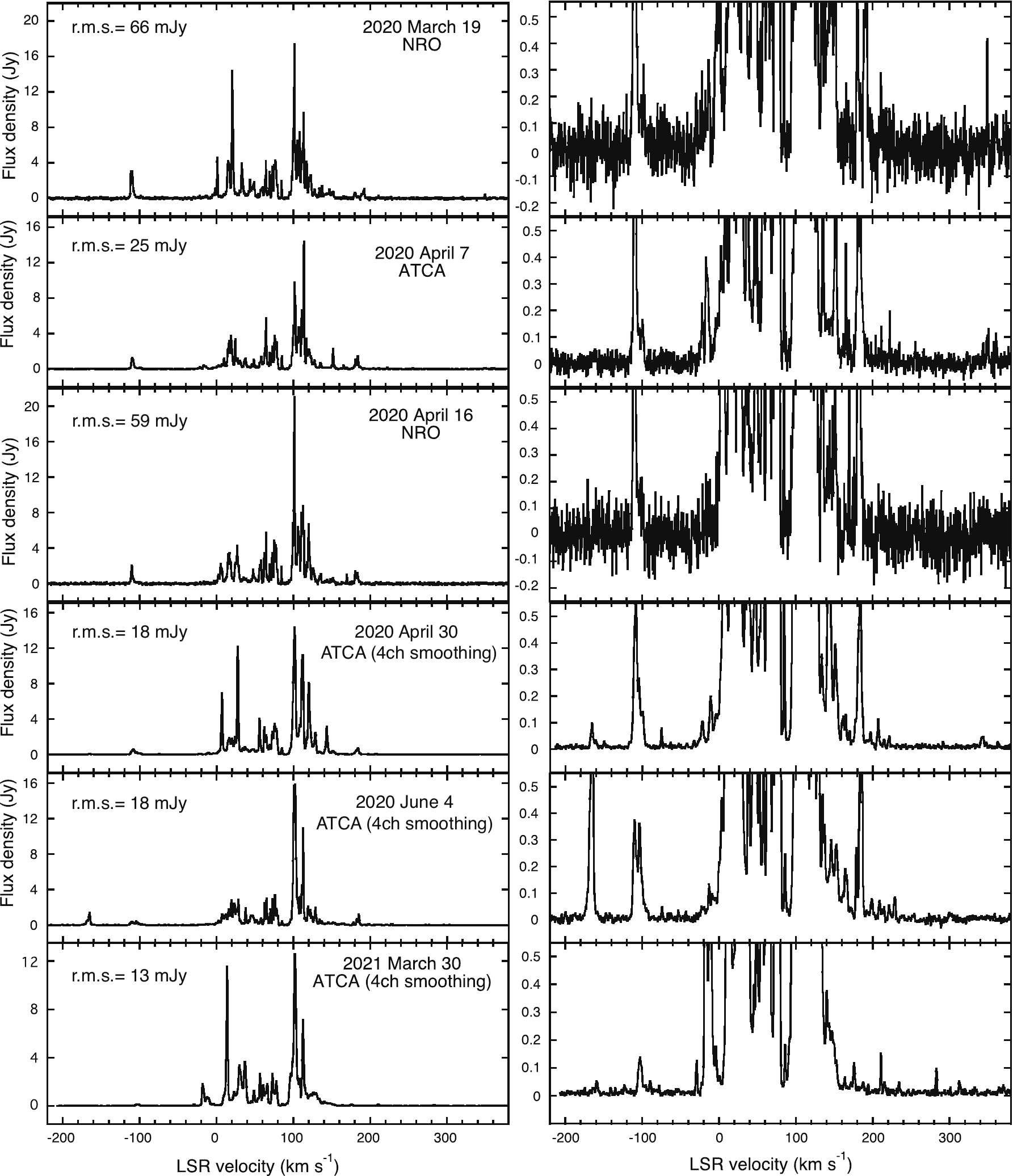} 
\caption{(Continued.)}
\end{figure*}

\begin{figure}
\begin{minipage}{8cm}
\includegraphics[width=8cm]{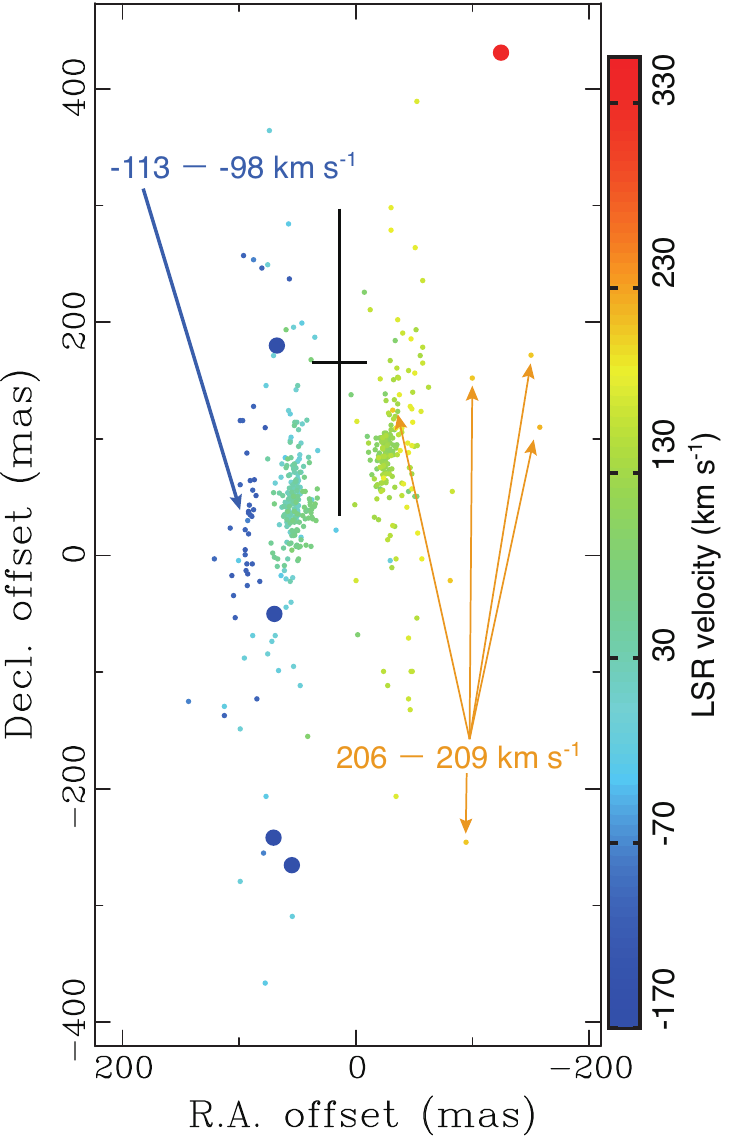}
\end{minipage}
\begin{minipage}{10cm}
\includegraphics[width=10cm]{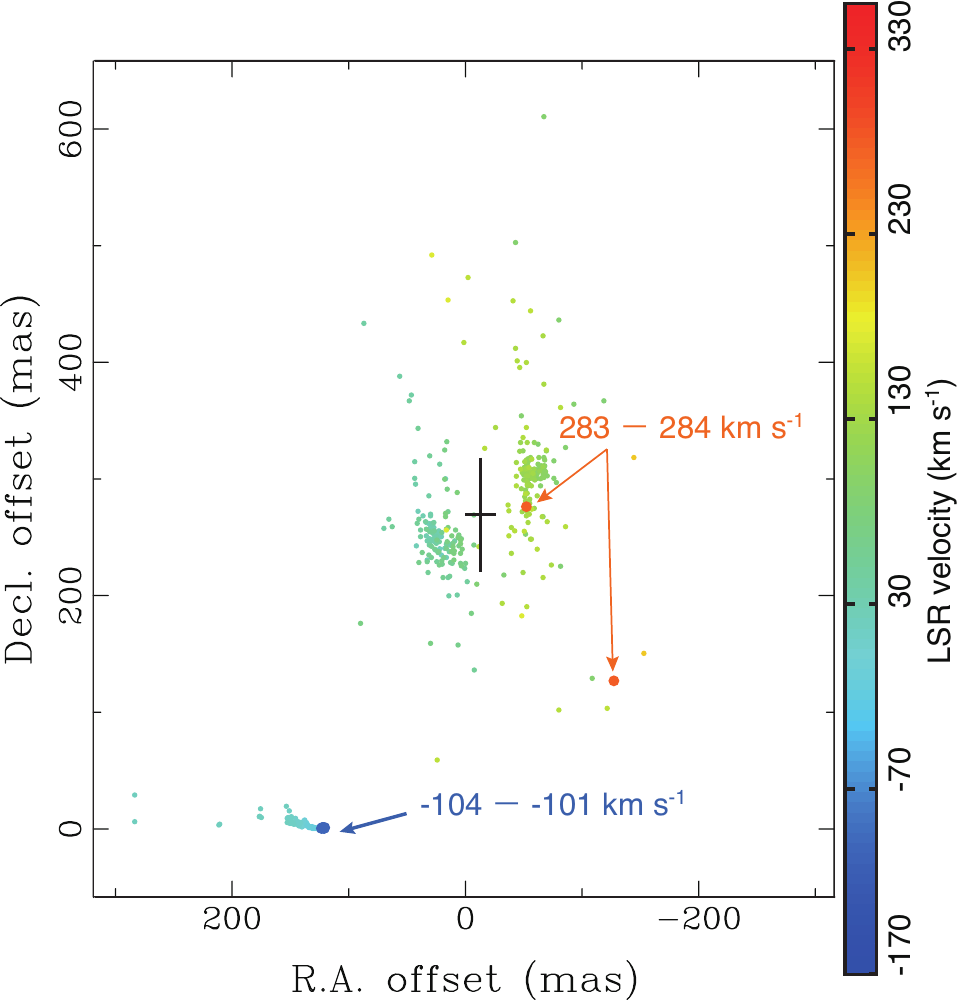}
\end{minipage}
\caption{{\it Left:} Map of H$_2$O masers in I18043 taken with ATCA on 2020 April 30. The maser distribution has 
a larger positional uncertainty in declination offset. 
The scatter of the weak (${\rm SNR}\simeq10$) blue-shifted ($V_{\rm LSR}=-113$--$-98$~km~s$^{-1}$) and the red-shifted ($V_{\rm LSR}=206$--209~km~s$^{-1}$) components,
is attributed to such a large positional uncertainty in this direction, but the each mean position of the blue- and red-shifted components are still close to the major alignment axis of the two major clusters of the masers.
The new blue-shifted components ($V_{\rm LSR}\sim -165$~km~s$^{-1}$) and the red-shifted components ($V_{\rm LSR}\sim$344~km~s$^{-1}$) are highlighted in bigger filled circles. The position and the size of a black cross indicate the position and statistical error (3$\sigma$) of the continuum source, respectively. The synthesized beam is 
\arcsa{2}{53}$\times$\arcsa{0}{34} at a position angle of $-$6.78\arcdeg.
{\it Right:} Same as the left panel but on 2021 March 30. The new blue-shifted components ($V_{\rm LSR}\sim -102$~km~s$^{-1}$) and the red-shifted components ($V_{\rm LSR}\sim$284~km~s$^{-1}$) are highlighted in bigger filled circles. The synthesized beam is
\arcsa{1}{42}$\times$\arcsa{0}{41} at a position angle of $-$1.27\arcdeg.
} \label{fig:ATCA-map}
\end{figure}

\begin{figure}
\begin{center}
\includegraphics[width=16cm]{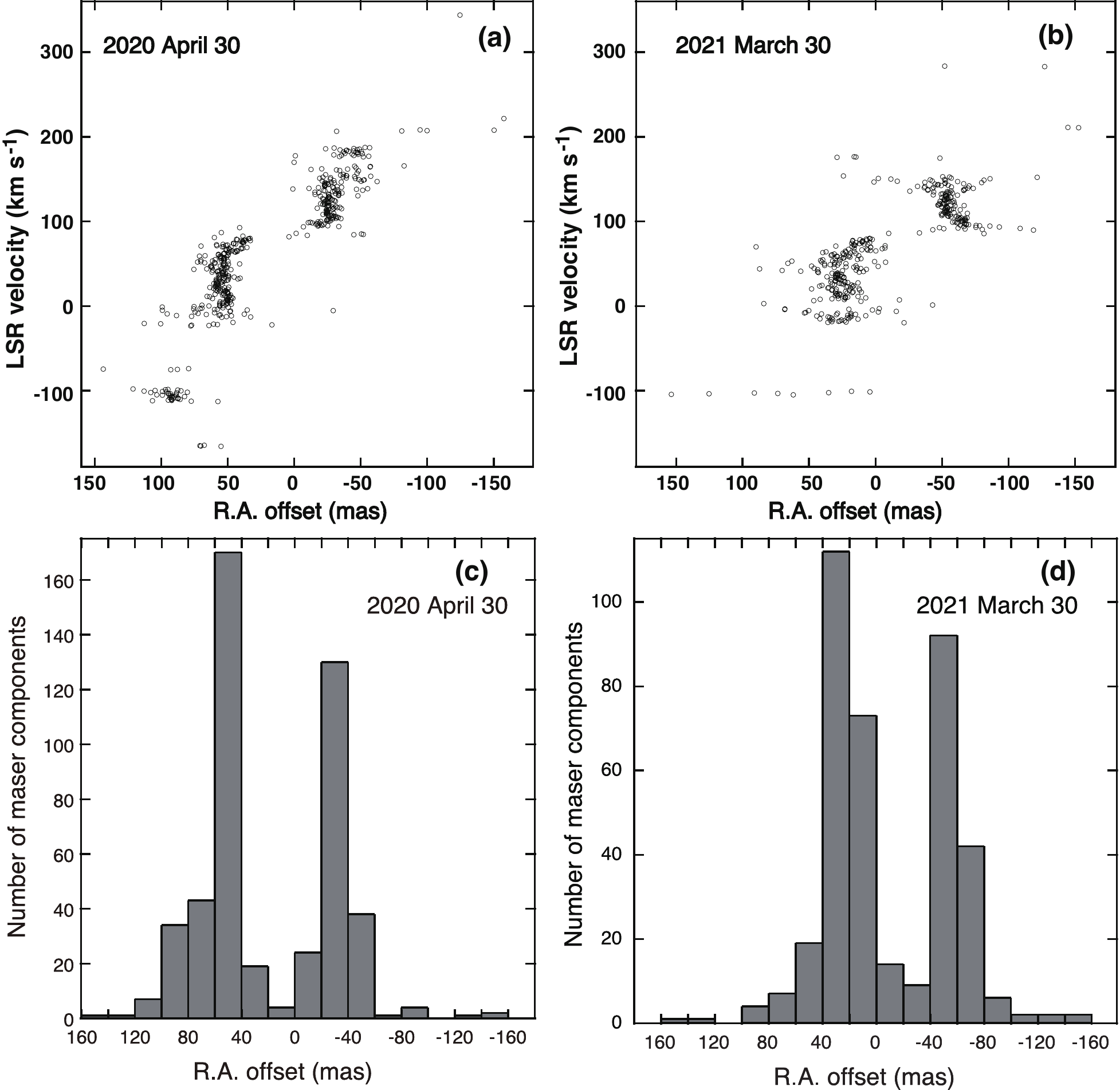}
\end{center}
\caption{Distribution of H$_2$O masers in I18043 along right ascension offset found with ATCA. (a) LSR velocity distribution of the masers on 2020 April 30. (b) Same as panel a, but on 2021 March 30. (c) Histogram of right ascension offsets on 30 April 2020. (d) Same as panel c, but on 30 March 2021.} \label{fig:ATCA-map2}
\end{figure}

\begin{figure*}
\begin{center}
\includegraphics[width=17cm]{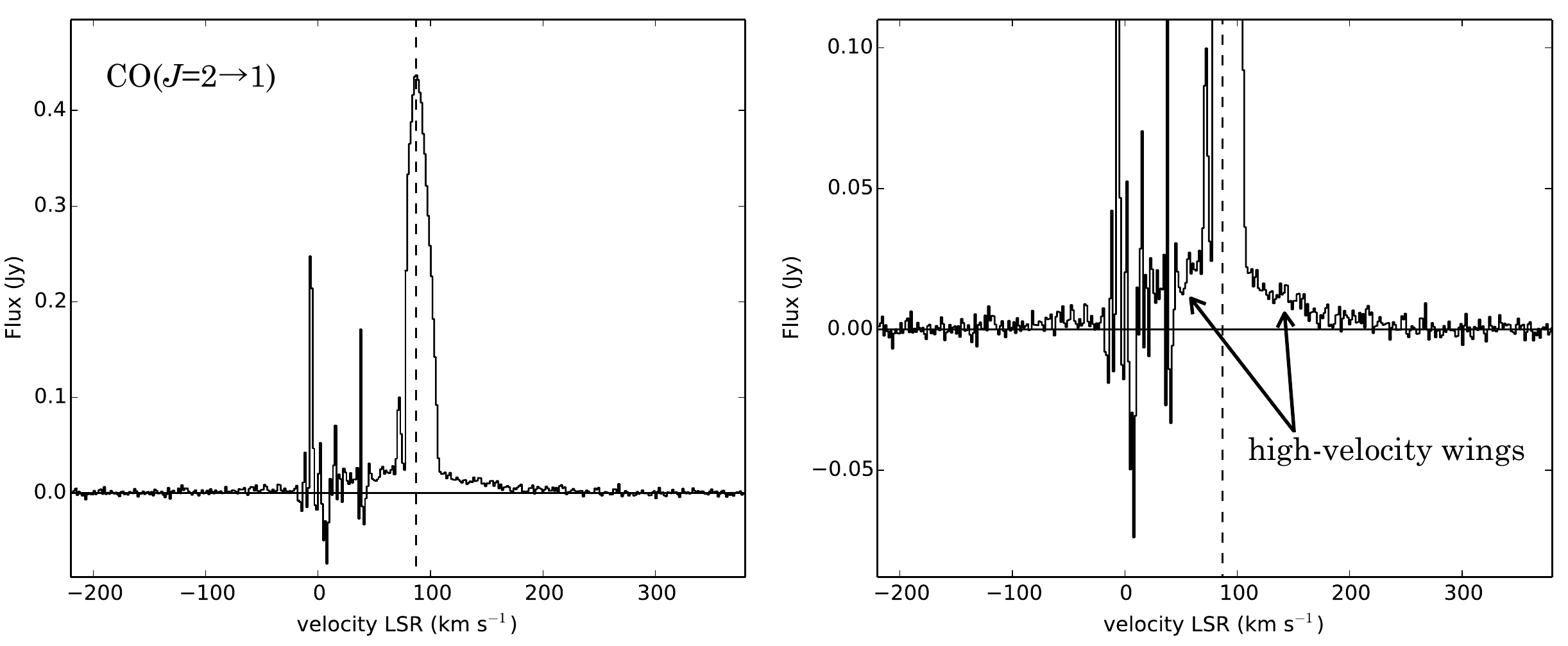} 
\end{center}
\caption{{\it Left:} ALMA observations of the CO($J$=2$\rightarrow$1) emission in I18043. The vertical dashed line indicates the systemic velocity of the source, $V_{\rm LSR, sys}=87$~km~s$^{-1}$. {\it Right:} Zoom-in toward the CO($J$=2$\rightarrow$1) emission around the systemic velocity to show the high-velocity wings.}
\label{fig:ALMA_CO2-1_spectrum}
\end{figure*}

\begin{figure*}
\begin{center}
\includegraphics[width=12cm]{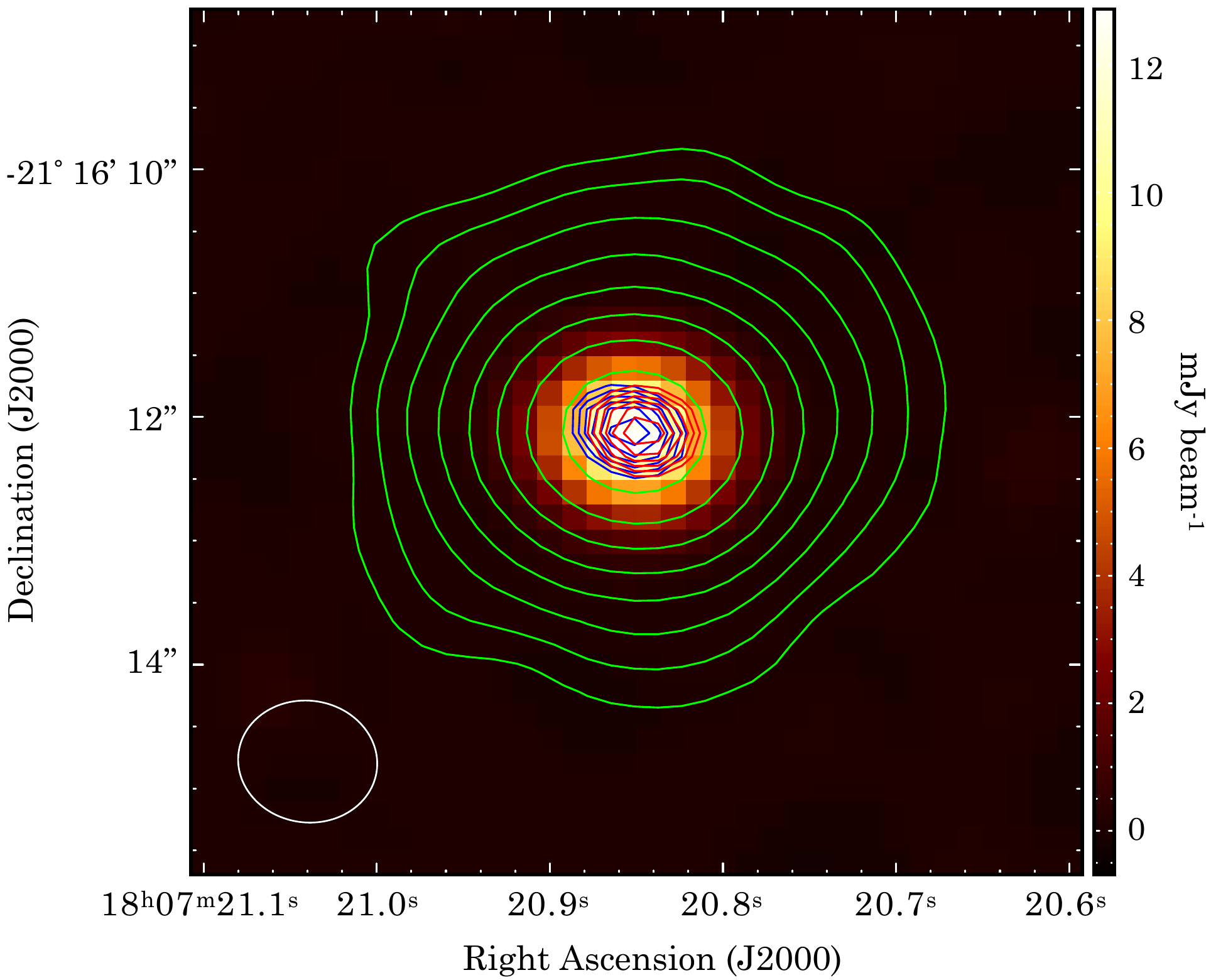} 
\end{center}
\caption{Integrated CO($J$=2$\rightarrow$1) emission around the systemic velocity of I18043 observed with ALMA. 
The green contours represent the velocity-integrated emission in the velocity range: 65--110~km~s$^{-1}$. The values of the contours are 3$\times$$\sigma$$\times$1.6$^{n}$, with $n$=0,1,2,3$\ldots$, and $\sigma$ is the rms value of the emission, 0.07~Jy~beam$^{-1}$~km~s$^{-1}$. The blue and red contours represent the velocity-integrated emission in the velocity ranges:$-$125--65~km~s$^{-1}$ and 110--300~km~s$^{-1}$, respectively. The values of the contours are 0.7, 0.75, 0.8, 0.85, 0.9 and 0.95 times the peak value of the emission, which is 0.62~Jy~beam$^{-1}$~km~s$^{-1}$ and 1.1~Jy~beam$^{-1}$~km~s$^{-1}$ for the blue- and red-shifted emission, respectively. The background image corresponds to the intensity of the CO($J$=2$\rightarrow$1) emission. See the scale to the right side. }
\label{fig:ALMA_CO2-1_map}
\end{figure*}

\begin{figure*}
\begin{center}
\includegraphics[width=17cm]{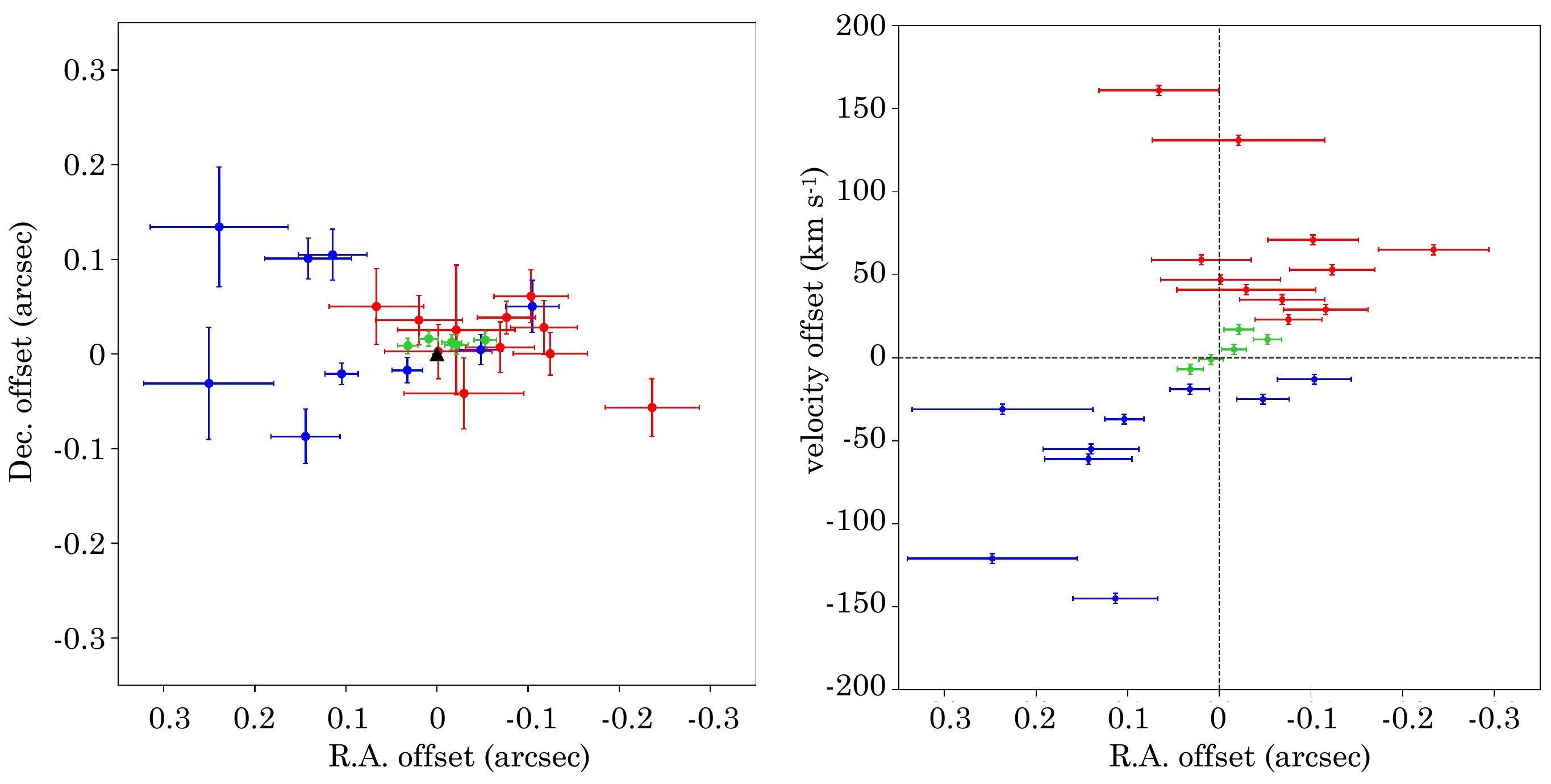} 
\end{center}
\caption{{\it Left}: Peak positions of the ALMA CO($J$=2$\rightarrow$1) emission in each bin toward I18043. The position of the 230~GHz continuum is indicated with a black triangle, R.A.(J2000)=18$^{\rm h}$07$^{\rm m}$20.851$^{\rm s}$, Dec.(J2000)=$-$21$^{\circ}$16$^{\prime}$12.13$^{\prime\prime}$. {\it Right}: Right Ascension offsets from the continuum peak position as a function of the velocity offset from the systemic velocity, $V_{\rm LSR, sys}=87$~km~s$^{-1}$. The blue, green and red marks represent emission in the velocity ranges: $-170<V_{\rm offset}<-$20~km~s$^{-1}$, $-10<V_{\rm offset}<$ 10~km~s$^{-1}$ and $10<V_{\rm offset}<$ 170~km~s$^{-1}$, respectively.
These figures seem to be obtained after averaging 2 spectral channels in the green velocity range and 9 spectral channels in the blue and red velocity ranges.
}\label{fig:ALMA_pos_PV}
\end{figure*}

\begin{figure*}
\begin{center}
\includegraphics[width=10cm]{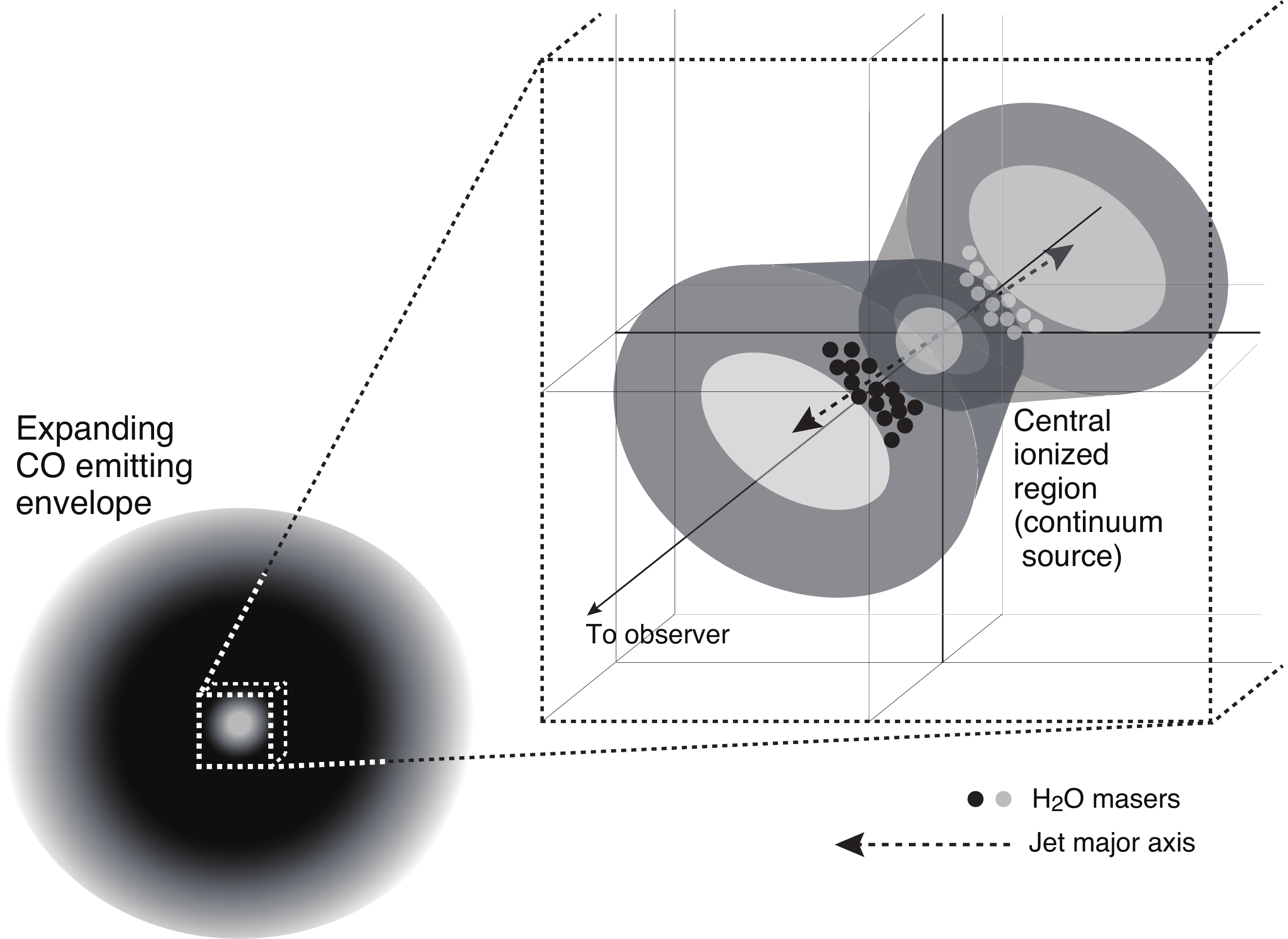} 
\end{center}
\caption{Schematic picture of the system of I18043. 
The whole expanding envelope (bottom-left corner) is traced in the 65–110 km s$^{-1}$ component of the CO emission (Figure \ref{fig:ALMA_CO2-1_map}). 
The envelope exhibits weak bipolarity in the east--west direction indicated by a velocity gradient (Figure \ref{fig:ALMA_pos_PV}). 
The center of the envelope is ionized (either by photons from the central star or by shocks at the outflow), as indicated by the 22 GHz continuum emission (Figure \ref{fig:ATCA-map}). 
This part is zoomed up in the right side, exhibiting a biconical outflow. 
The distribution of H$_2$O masers (black and grey circles for the blue- and red-shifted components, respectively) exhibits 
strong bipolarity in the present ATCA (Figure \ref{fig:ATCA-map}) and VLBA \citep{oro17} data. 
}\label{fig:schematic-picture}
\end{figure*}

\end{document}